# Duality for Several Families of Evaluation Codes

Maria Bras-Amorós *and Michael E. O'Sullivan†

October 14, 2018

**Abstract**

We consider generalizations of Reed-Muller codes, toric codes, and codes from certain plane curves, such as those defined by norm and trace functions on finite fields. In each case we are interested in codes defined by evaluating arbitrary subsets of monomials, and in identifying when the dual codes are also obtained by evaluating monomials. We then move to the context of order domain theory, in which the subsets of monomials can be chosen to optimize decoding performance using the Berlekamp-Massey-Sakata algorithm with majority voting. We show that for the codes under consideration these subsets are well-behaved and the dual codes are also defined by monomials.

## 1 Introduction

Goppa's introduction of algebraic geometry into coding theory led to several methods for constructing codes. One prominent and very general family is codes from order domains [9, 14], which include Reed-Solomon codes, Reed-Muller codes, one-point codes, and certain codes from higher dimensional algebraic varieties [13]. For these codes there is a lower bound on the minimum distance and an efficient generalization of the Berlekamp-Massey algorithm based on Sakata's algorithm for finding linear recurrence relations in multidimensional arrays [17]. Using the majority voting method of Feng, Rao, and Duursma, [4, 3] the algorithm corrects up to half the above mentioned bound on the minimum distance.

Improvements to code dimension, while maintaining a fixed decoding radius were discovered by Feng and Rao [4]. Our earlier work [2] considered a different improvement, based on the observation that the decoding algorithm corrects an error vector based not so much on the weight of the vector but rather the "footprint" of the error locations [6]. For some codes, most error vectors of a particular weight $t$ are correctable with fewer check symbols than are required to correct all vectors of weight $t$. These results may be combined with the improvement to dimension due to Feng and Rao.

*Universitat Autònoma de Barcelona, mbras@deic.uab.cat
†San Diego State University, mosulliv@sciences.sdsu.edu

In the discussion above, the order domain is used to construct the check matrix, so an important question for practical implementation is to identify a generating matrix for the code. In this paper we consider several order domains and parity-check matrices constructed by evaluating monomials from the order domain at points on the variety determined by the order domain. We begin by investigating duality for these codes, independently of the correction capability issues. More precisely, we first identify minimal generating sets for codes generated by monomials. We then derive conditions ensuring that the dual code to a code generated by monomials is also generated by monomials, and we find the generators for the dual code in this case. Sections 2 and 3 treat generalizations of Reed-Muller codes and toric codes, respectively. Section 4 covers codes from algebraic curves with structure similar to Hermitian curves. Finally, we turn to codes with prescribed error correction capability. Section 5 introduces order domains, summarizes the results on improving dimension, and treats the application of these results to the codes of Sections 2-4.

# 2 Reed-Muller codes

Let $n = q^m$ and call $P_1, \ldots, P_n$ the $n$ points in $\mathbb{F}_q^m$. The *Reed-Muller original code* $RM_q(s, m)$ is defined as the image of the map $\varphi_s : \mathbb{F}_q[x_1, \ldots, x_m]_{<s} \longrightarrow \mathbb{F}_q^n$, $f \longmapsto (f(P_1), \ldots, f(P_n))$, where $\mathbb{F}_q[x_1, \ldots, x_m]_{<s}$ is the subspace of $\mathbb{F}_q[x_1, \ldots, x_m]$ of polynomials with total degree $< s$.

Let $A = \mathbb{F}_q[x_1, \ldots, x_m]$. Consider $\ll$ the graded lexicographic order on monomials in $A$ and let $z_i$ be the $i$th monomial in $A$ with respect to $\ll$, starting with $z_0 = 1$.

Let $\varphi : A \longrightarrow \mathbb{F}_q^n$ be such that $f \mapsto (f(P_1), \ldots, f(P_n))$ and let $W$ be a set of monomials in $A$. We call $E_W$ the vector space spanned by $\{\varphi(z_i) : z_i \in W\}$ and say that it is an *evaluation code*. Its dual code is denoted $C_W$. Given $W$ define $W^\varphi$ as the set $\{z_i \in W : \varphi(z_i)$ is not in the span of $\varphi(z_j), j \in W, j < i\}$. The interest in this set is that $E_{W^\varphi} = E_W$ while $W^\varphi$ might be strictly included in $W$. We say that a set of monomials $W$ is *divisor-closed* if $z_i \in W$ whenever $z_i$ divides a monomial in $W$.

Through all this section we will use the notation $\mathcal{M}$ for the set of monomials $x_1^{a_1} \cdots x_m^{a_m}$ such that $a_l < q$ for all $l$.

## 2.1 Dropping redundant parity checks

**Lemma 2.1.**

**(i)** *For each monomial* $z_i$ *in* $A$ *there exists a monomial* $z_{i'}$ *in* $\mathcal{M}$ *with* $z_{i'} \mid z_i$ *and* $\varphi(z_{i'}) = \varphi(z_i)$.

**(ii)** $\{\varphi(z_i) : z_i \in \mathcal{M}\}$ *is a basis of* $\mathbb{F}_q^n$.

*Proof.*

**(i)** If $z_i = x_1^{a_1} \cdots x_m^{a_m}$ take $z_{i'} = x_1^{b_1} \cdots x_m^{b_m}$ with

$$b_l = \begin{cases} q-1 & \text{if } q-1 \text{ divides } a_l \text{ and } a_l \neq 0, \\ a_l \bmod q-1 & \text{otherwise.} \end{cases}$$

Then $z_{i'} \in \mathcal{M}$ and it satisfies $z_{i'} \mid z_i$ and $\varphi(z_{i'}) = \varphi(z_i)$.

**(ii)** Since $\varphi$ is surjective and any polynomial in $A$ is a sum of monomials, by (i) the vectors in $\{\varphi(z_i) : z_i \in \mathcal{M}\}$ generate $\mathbb{F}_q^n$. Now, since the number of vectors in the set is at most $n$, they must constitute a basis of $\mathbb{F}_q^n$.

□

**Proposition 2.2.**

- *If $W \subseteq \mathcal{M}$ then $W^\varphi = W$.*
- *If $W$ is divisor-closed then $W^\varphi = W \cap \mathcal{M}$.*

*Proof.* The first item is a consequence of Lemma 2.1(ii).

For the second item suppose $z_i \in W$. If $z_i \in \mathcal{M}$ then, by Lemma 2.1, $\varphi(z_i)$ is linearly independent of all vectors $\varphi(z_j)$ with $z_i$ not dividing $z_j$. Since $z_i \mid z_j$ implies $i \leqslant j$, $\varphi(z_i)$ is not in the span of $\varphi(z_j), j \in W, j < i$. Hence, $z_i \in W^\varphi$. If $z_i \notin \mathcal{M}$ then there exists $z_{i'} \in \mathcal{M}$ with $z_{i'} \mid z_i$ and $\varphi(z_{i'}) = \varphi(z_i)$. Since $W$ is divisor-closed, we have $z_{i'} \in W$ and $i' < i$. Hence, $z_i \notin W^\varphi$. □

## 2.2 Dual codes

**Lemma 2.3.** *Suppose $x_1^{a_1} \cdots x_m^{a_m} \in \mathcal{M}$ and let $Z$ be the set of indices $l$ such that $a_l = q-1$. Then $\varphi(x_1^{a_1} \cdots x_m^{a_m}) \cdot \varphi(x_1^{b_1} \cdots x_m^{b_m}) \neq 0$ with $x_1^{b_1} \cdots x_m^{b_m} \in \mathcal{M}$ if and only if $b_l = q-1-a_l$ for all $l \notin Z$ and $b_l \in \{0, q-1\}$ for all $l \in Z$.*

*Proof.* Notice that by the distributive law

$$\varphi(x_1^{a_1} \cdots x_m^{a_m}) \cdot \varphi(x_1^{b_1} \cdots x_m^{b_m}) = \sum_{\alpha_1 \in \mathbb{F}_q} \alpha_1^{a_1+b_1} \sum_{\alpha_2 \in \mathbb{F}_q} \alpha_2^{a_2+b_2} \cdots \sum_{\alpha_m \in \mathbb{F}_q} \alpha_m^{a_m+b_m}.$$

From this formula it follows that $\varphi(x_1^{a_1} \cdots x_m^{a_m}) \cdot \varphi(x_1^{b_1} \cdots x_m^{b_m}) \neq 0$ if and only if, for all $l$, $a_l + b_l$ is $k_l(q-1)$ for some integer $k_l > 0$. Since $a_l, b_l < q$ this is only possible for $k_l = 1$ and $k_l = 2$. This is equivalent to have either $a_l = b_l = q-1$ or $b_l = q-1-a_l$. □

**Proposition 2.4.** *If $W \in \mathcal{M}$ is divisor-closed then $C_W = E_{W^\perp}$ where $W^\perp = \mathcal{M} \setminus \{x_1^{q-1-a_1} \cdots x_m^{q-1-a_m} : x_1^{a_1} \cdots x_m^{a_m} \in W\}$.*

*Proof.* By Lemma 2.1(ii), the dimension of $C_W$ is equal to the dimension of $E_{W^\perp}$. Hence it is enough to prove that $\varphi(x_1^{a_1} \cdots x_m^{a_m}) \cdot \varphi(x_1^{b_1} \cdots x_m^{b_m}) = 0$ for all $x_1^{a_1} \cdots x_m^{a_m} \in W$ and all $x_1^{b_1} \cdots x_m^{b_m} \in W^\perp$.

First of all notice that, since $W$ is divisor-closed, if $x_1^{a_1}\cdots x_m^{a_m} \in W$ then any monomial $x_1^{a'_1}\cdots x_m^{a'_m}$ with $a'_l \in \{0, a_l\}$ for all $l$ is also in $W$. Thus no monomial $x_1^{b'_1}\cdots x_m^{b'_m}$ with $b'_l \in \{q-1, q-1-a_l\}$ for all $l$ is in $W^\perp$.

Now, let $Z$ be the set of indices $l$ such that $a_l = q-1$. By Lemma 2.3, $\varphi(x_1^{a_1}\cdots x_m^{a_m})\cdot\varphi(x_1^{b_1}\cdots x_m^{b_m}) \neq 0$ if and only if $b_l = q-1-a_l$ for all $l \notin Z$ and $b_l \in \{0, q-1\}$ for all $l \in Z$. But then, by the previous argument, $x_1^{b_1}\cdots x_m^{b_m} \notin W^\perp$. □

**Remark 2.5.** The condition of $W$ being divisor-closed in the previous proposition is much stronger than what we really need. Indeed, it would be enough to have $W$ satisfying that for any $x_1^{a_1}\cdots x_m^{a_m} \in W$ and $Z_a$ being the set of indices $l$ with $a_l = q-1$, any monomial $x_1^{a'_1}\cdots x_m^{a'_m}$ is also in $W$ where $a'_l \in \{0, q-1\}$ for all $l \in Z_a$ and $a'_l = a_l$ otherwise.

# 3 Toric codes

Let $A = \mathbb{F}_q[x_1, \ldots, x_m]$ and $\{z_i : i \in \mathbb{N}_0\}$ be as before. Now let $n = (q-1)^m$ and let $P_1, \ldots, P_n$ be the $n$ points in $(\mathbb{F}_q^*)^m$. Let $\varphi : A \longrightarrow \mathbb{F}_q^n$ be such that $f \mapsto (f(P_1), \ldots, f(P_n))$. Given a set of monomials $W$ we call $E_W$ the vector space spanned by $\{\varphi(z_i) : z_i \in W\}$ and $C_W$ its orthogonal space. Let $Q$ be an integral convex polytope such that the intersection $Q \cap \mathbb{Z}^m$ is properly contained in $[0 \ldots q-2]^m$. Let $W$ be the set of monomials $x_1^{a_1}\cdots x_m^{a_m}$ in $A$ whose exponents $(a_1, \ldots, a_m)$ correspond to points in $Q \cap \mathbb{Z}^m$. The code $E_W$ is called a *toric code* [7, 8, 12].

For toric codes we will use $\mathcal{M}$ for the set of monomials $x_1^{a_1}\cdots x_m^{a_m}$ such that $a_l < q-1$ for all $l$.

## 3.1 Dropping redundant parity checks

**Lemma 3.1.**

**(i)** *For each monomial $z_i$ in $A$ there exists a monomial $z_{i'}$ in $\mathcal{M}$ with $z_{i'} \mid z_i$ and $\varphi(z_{i'}) = \varphi(z_i)$.*

**(ii)** *$\{\varphi(z_i) : z_i \in \mathcal{M}\}$ is a basis of $\mathbb{F}_q^n$.*

**(iii)** *$\{\varphi((x_1\cdots x_m)z_i) : z_i \in \mathcal{M}\}$ is a basis of $\mathbb{F}_q^n$.*

*Proof.*

**(i)** If $z_i = x_1^{a_1}\cdots x_m^{a_m}$ take $z_{i'} = x_1^{b_1}\cdots x_m^{b_m}$ with $b_l = a_l \bmod q-1$ for all $l$. Then $z_{i'} \in \mathcal{M}$ and it satisfies $z_{i'} \mid z_i$ and $\varphi(z_{i'}) = \varphi(z_i)$.

**(ii)** Since $\varphi$ is surjective and any polynomial in $A$ is a sum of monomials, by (i) the vectors in $\{\varphi(z_i) : z_i \in \mathcal{M}\}$ generate $\mathbb{F}_q^n$. Now, since the number of vectors in the set is at most $n$, they must constitute a basis of $\mathbb{F}_q^n$.

**(iii)** By (ii) it is enough to prove that for any $x_1^{a_1} \cdots x_m^{a_m} \in \mathcal{M}$ there exists $x_1^{b_1} \cdots x_m^{b_m} \in \mathcal{M}$ such that $\varphi(x_1^{a_1} \cdots x_m^{a_m}) = \varphi(x_1 \cdots x_m \cdot x_1^{b_1} \cdots x_m^{b_m}) = \varphi(x_1^{b_1+1} \cdots x_m^{b_m+1})$. Take

$$b_l = \begin{cases} a_l - 1 & \text{if } a_l > 0, \\ a_l = q - 2 & \text{otherwise.} \end{cases}$$

It is easy to verify that the previous equality holds.

□

**Proposition 3.2.** *Given a toric code $E_W$ there is no proper subset $W'$ of $W$ such that $E_{W'} = E_W$.*

*Proof.* It is a consequence of Lemma 3.1(i) and the fact that, by definition, $W \subseteq \mathcal{M}$. □

## 3.2 Dual codes

From now on we will consider $\psi$ as the map $A \longrightarrow \mathbb{F}_q^n$ such that $f \mapsto \varphi(f \cdot x_1 \cdots x_m)$.

**Lemma 3.3.** *Given $x_1^{a_1} \cdots x_m^{a_m} \in \mathcal{M}$, $\varphi(x_1^{a_1} \cdots x_m^{a_m}) \cdot \psi(x_1^{b_1} \cdots x_m^{b_m}) \neq 0$ with $x_1^{b_1} \cdots x_m^{b_m} \in \mathcal{M}$ if and only if $b_l = q - 2 - a_l$ for all $l$.*

*Proof.* Notice that

$$\varphi(x_1^{a_1} \cdots x_m^{a_m}) \cdot \psi(x_1^{b_1} \cdots x_m^{b_m}) = \sum_{\alpha_1 \in \mathbb{F}_q^*} \alpha_1^{a_1+b_1+1} \sum_{\alpha_2 \in \mathbb{F}_q^*} \alpha_2^{a_2+b_2+1} \cdots \sum_{\alpha_m \in \mathbb{F}_q^*} \alpha_m^{a_m+b_m+1}.$$

From this formula it follows that $\varphi(x_1^{a_1} \cdots x_m^{a_m}) \cdot \psi(x_1^{b_1} \cdots x_m^{b_m}) \neq 0$ if and only if, for all $l$, $a_l + b_l + 1$ is $k_l(q-1)$ for some integer $k_l > 0$. Since $a_l, b_l < q - 1$ this is only possible for $k_l = 1$. Thus, the former product is non-zero if and only if $a_l + b_l$ is $q - 2$ for all $l$. □

**Definition 3.4.** *Given $z = x_1^{a_1} \cdots x_m^{a_m}$ define $\hat{z} = x_1^{b_1} \cdots x_m^{b_m}$ where*

$$b_l = \begin{cases} q - 1 - a_l & \text{if } a_l > 0, \\ 0 & \text{otherwise.} \end{cases}$$

**Proposition 3.5.** *Given any toric code $E_W$, let $W^\perp = \mathcal{M} \setminus \{\hat{z} : z \in W\}$. Then $C_W = E_{W^\perp}$.*

*Proof.* Let $W' = \mathcal{M} \setminus \{x_1^{q-2-a_1} \cdots x_m^{q-2-a_m} : x_1^{a_1} \cdots x_m^{a_m} \in W\}$. Then $C_W$ is the vector space spanned by $\psi(W')$. Indeed, by Lemma 3.3 the vector space spanned by $\psi(W')$ is in $C_W$. By Lemma 3.1, the vector space spanned by $\psi(W')$ has dimension $n - \dim(E_W)$. Thus it is $C_W$. Now it is easy to check that the vector space spanned by $\psi(W')$ is exactly $E_{W^\perp}$. □

This result was obtained independently in [16, 15].

**Remark 3.6.** $C_W$ might not be a toric code. Indeed, it will be a toric code if and only if $W^\perp$ is the set of monomials $x_1^{a_1}\cdots x_m^{a_m}$ whose exponents $(a_1,\dots,a_m)$ correspond to points in $Q\cap\mathbb{Z}^m$, where $Q$ is a polytope such that $Q\cap\mathbb{Z}^m$ is properly contained in $[0\dots q-2]^m$.

# 4 Norm-trace codes and generalizations

Let $q$ be a prime power and $r$ an integer greater than or equal to 2. The curve defined over $\mathbb{F}_{q^r}$ with affine equation

$$x^{\frac{q^r-1}{q-1}} = y^{q^{r-1}} + y^{q^{r-2}} + \cdots + y.$$

is called the *norm-trace curve* associated to $q$ and $r$. In fact, the defining equation is equivalent to

$$\mathcal{N}_{\mathbb{F}_{q^r}/\mathbb{F}_q}(x) = \mathcal{T}_{\mathbb{F}_{q^r}/\mathbb{F}_q}(y),$$

where, for $x\in\mathbb{F}_{q^r}$, $\mathcal{N}_{\mathbb{F}_{q^r}/\mathbb{F}_q}(x)$ denotes the norm of $x$ over $\mathbb{F}_q$, and for $y\in\mathbb{F}_{q^r}$, $\mathcal{T}_{\mathbb{F}_{q^r}/\mathbb{F}_q}(y)$ denotes the trace of $y$ over $\mathbb{F}_q$. Norm-trace curves were studied by Geil in [5]. They are a natural generalization of Hermitian curves, these being norm-trace curves resulting from the field extension $\mathbb{F}_{q^2}/\mathbb{F}_q$. Norm-trace curves have a single rational point $P_\infty$ at infinity and $n = q^{2r-1}$ proper rational points.

In this section we consider a somewhat broader family of curves that include norm-trace curves. For example, our family includes $x^u = \mathcal{T}_{\mathbb{F}_{q^r}/\mathbb{F}_q}(y)$, where $u$ divides $(q^r-1)/(q-1)$, and it also includes the maximal curves derived from Hermitian curves studied in [10].

A linearized polynomial over $\mathbb{F}_q$—also called a $q$ polynomial—is a polynomial over $\mathbb{F}_q$ whose terms all have degree a power of $q$ [11, §3.4]. Let $L(y) = \sum_{i=0}^d a_i y^{q^i}$ be a linearized polynomial such that $a_0, a_d$ are nonzero and such that $L(y)=0$ has $q^d$ distinct solutions in $\mathbb{F}_{q^r}$. Then $L$ gives a $q^d$-to-one map from $\mathbb{F}_{q^r}$ into itself [11].

Let $\eta$ be a primitive element of $\mathbb{F}_{q^r}$. Let $v$ be any factor of $q^r-1$ and let $D$ be the powers of $\eta^v$, along with 0: $D=\{0\}\cup\{\eta^{vm} : m\in\{1,\dots,\frac{q^r-1}{v}\}\}$. We will assume that $L(\mathbb{F}_{q^r})\supseteq D$. Then for any $u$ dividing $v$ we consider the curve $x^u = L(y)$ whose coordinate ring is

$$A = \frac{\mathbb{F}_{q^r}[x,y]}{x^u - L(y)}.$$

A basis of $A$ as a vector space over $\mathbb{F}_{q^r}$ is given by the images in $A$ of the set of monomials $\mathcal{B} = \{x^a y^b : b < q^d\}$. Consider $\ll$ the total ordering on $A$ determined by the $(q^d,u)$ weighted degree, $\deg_{q^d,u} x^a y^b = q^d a + ub$. One can check that any monomial in $\mathbb{F}_q[x,y]$ has the same weighted degree as exactly one monomial in $\mathcal{B}$. In particular, no two monomials from $\mathcal{B}$ have the same weighted degree. Let $z_i$ be the $i$th monomial in $\mathcal{B}$ with respect to $\ll$, starting with $z_0 = 1$. Notice that if $z_i$ divides $z_j$ then $i\leqslant j$. We say that a set of monomials $W$ is divisor-closed if $z_i\in W$ whenever $z_i$ divides a monomial in $W$.

Since $u$ is coprime to $q$, this curve has a unique point at infinity. We consider all points $(\alpha, \beta)$ on the curve such that $\alpha^u = L(\beta) \in D$. Thus we have $\alpha$ is either 0 or $\alpha \in \{\eta^{mv/u} : m \in \{1, \ldots, \frac{(q^r-1)u}{v}\}\}$. There are $n = q^d(\frac{(q^r-1)u}{v} + 1)$ such points, $P_1, \ldots, P_n$.

Let $\varphi : A \longrightarrow \mathbb{F}_q^n$ be such that $f \mapsto (f(P_1), \ldots, f(P_n))$ and let $W$ be a set of monomials in $\mathcal{B}$. We call $E_W$ the vector space spanned by $\{\varphi(z_i) : z_i \in W\}$ and $C_W$ the dual code.

## 4.1 Dropping redundant parity checks

As in the previous section, given $W$ define $W^\varphi$ as the set $\{z_i \in W : \varphi(z_i)$ is not in the span of $\varphi(z_j), j \in W, j < i\}$. Let $\mathcal{M} = \{x^a y^b : a \leqslant \frac{(q^r-1)u}{v}, b < q^d\}$.

**Lemma 4.1.**

**(i)** *For each monomial $x^a y^b \in \mathcal{B}$ there exists a monomial $x^{a'} y^{b'}$ in $\mathcal{M}$ with $x^{a'} y^{b'} \mid x^a y^b$ and $\varphi(x^{a'} y^{b'}) = \varphi(x^a y^b)$.*

**(ii)** *$\{\varphi(x^a y^b) : x^a y^b \in \mathcal{M}\}$ is a basis of $\mathbb{F}_q^n$.*

*Proof.*

**(i)** Suppose $x^a y^b \in \mathcal{B}$, with $b < q^d$ and let $b' = b$ and

$$a' = \begin{cases} \frac{(q^r-1)u}{v} & \text{if } \frac{(q^r-1)u}{v} \text{ divides } a \text{ and } a \neq 0, \\ a \mod \frac{(q^r-1)u}{v} & \text{otherwise.} \end{cases}$$

Then $x^{a'} y^{b'}$ is in $\mathcal{M}$ and it satisfies $x^{a'} y^{b'} \mid x^a y^b$ and $\varphi(x^{a'} y^{b'}) = \varphi(x^a y^b)$.

**(ii)** Since $\varphi$ is surjective and any polynomial in $A$ is a sum of monomials in $\mathcal{B}$, by (i) the vectors in $\{\varphi(z_i) : z_i \in \mathcal{M}\}$ generate $\mathbb{F}_q^n$. Now, since the number of vectors in the set is at most $n$, they must constitute a basis of $\mathbb{F}_q^n$.

□

**Proposition 4.2.**

- *If $W \subseteq \mathcal{M}$ then $W^\varphi = W$.*
- *If $W$ is divisor-closed then $W^\varphi = W \cap \mathcal{M}$.*

*Proof.* Analogous to the proof of Proposition 2.2. □

## 4.2 Dual codes

In this section we make the additional assumption that the prime divisor of $q$ also divides $v - u$.

**Proposition 4.3.** *For $x^{a_1}y^{b_1}$ and $x^{a_2}y^{b_2}$ in $\mathcal{M}$, $\varphi(x^{a_1}y^{b_1})\cdot\varphi(x^{a_2}y^{b_2}) = 0$ unless $a_1 + a_2$ is $\frac{(q^r-1)u}{v}$ or $2\frac{(q^r-1)u}{v}$ and $b_1 + b_2$ is in $\{2q^d - q^l - 1 : l \in \{0, \dots, d\}\}$.*

*Proof.* Let $a = a_1 + a_2$ and $b = b_1 + b_2$ and notice that $b \leqslant 2(q^d - 1)$. By the distributive law

$$\varphi(x^{a_1}y^{b_1}) \cdot \varphi(x^{a_2}y^{b_2}) = \sum_{c\in D} \left( \sum_{\substack{\alpha\in\mathbb{F}_{q^r} \\ \alpha^u=c}} \alpha^a \right) \left( \sum_{\substack{\beta\in\mathbb{F}_{q^r} \\ L(\beta)=c}} \beta^b \right)$$

We prove in Lemma 4.6 that the second term is independent of $c$, for $b \leqslant 2(q^d - 1)$, so it may be taken outside the summation over $c$. We also show that it is nonzero only if $b \in \{2q^d - q^l - 1 : l \in \{0, \dots, d\}\}$. In Lemma 4.7, we show the first term is 0 unless $a$ is a positive multiple of $\frac{(q^r-1)u}{v}$. This gives the result. □

This is exactly what we need to establish the dual code when $W$ is divisor-closed.

**Proposition 4.4.** *If $W \in \mathcal{M}$ is divisor-closed then $C_W = E_{W^\perp}$ where*

$$W^\perp = \mathcal{M} \setminus \{x^{\frac{(q^r-1)u}{v}-a} y^{q^d-1-b} : x^a y^b \in W\}.$$

*Proof.* By Lemma 4.1(ii), the dimension of $C_W$ is equal to the dimension of $E_{W^\perp}$. Hence it is enough to prove that $\varphi(x^{a_1}y^{b_1})\varphi(x^{a_2}y^{b_2}) = 0$ for all $x^{a_1}y^{b_1} \in W$ and all $x^{a_2}y^{b_2} \in W^\perp$.

First of all notice that, since $W$ is divisor-closed, if $x^{a_1}y^{b_1} \in W$ then any monomial $x^{a_1'}y^{b_1'}$ with $a_1' \leqslant a_1$ and any $b_1' \leqslant b_1$ is also in $W$. Thus no monomial $x^{a_2'}y^{b_2'}$ with $a_2' \geqslant \frac{(q^r-1)u}{v} - a$ and $b_2' \geqslant q^d - 1 - b_1$ is in $W^\perp$.

Now, from the previous proposition, if $\varphi(x^{a_1}y^{b_1})\varphi(x^{a_2}y^{b_2}) \neq 0$ then $a_1+a_2 \geqslant \frac{(q^r-1)u}{v}$ and $b_1 + b_2 \geqslant q^d - 1$. Thus $x^{a_2}y^{b_2} \notin W^\perp$. □

**Remark 4.5.** The condition of $W$ being divisor-closed in the previous proposition is much stronger than what we really need. Indeed, it would be enough to have $W$ satisfying that for any $x^a y^b \in W$ the monomials $x^{a'}y^{b'}$ are also in $W$ where $a' \in \{a, a - \frac{(q^r-1)u}{v}\} \cap \mathbb{N}_0$ and $b' \in \{b - q^d + q^l : l \in \{0 \dots d\}\} \cap \mathbb{N}_0$.

We now establish the claims made in the proof of Proposition 4.3. The proof of the following Lemma is based on a proof for the trace function from $\mathbb{F}_{q^r}$ to $\mathbb{F}_q$ in [5].

**Lemma 4.6.** *Let $L(x)$ be a q-polynomial of degree $q^d$ and nonzero linear term, with all its roots in $\mathbb{F}_{q^r}$. For any $b$ satisfying $0 \leqslant b \leqslant 2(q^d - 1)$, $\sum_{L(\beta)=c} \beta^b$ is independent of $c$ in the image of $L(x)$. Furthermore it can be nonzero only when $b \in \{2q^d - q^l - 1 : l \in \{0, \dots, d\}\}$.*

*Proof.* Let $p$ be the prime divisor of $q$. Since

$$\left(\sum_{\substack{\beta\in\mathbb{F}_{q^r}\\ L(\beta)=c}} \beta^b\right)^p = \sum_{\substack{\beta\in\mathbb{F}_{q^r}\\ L(\beta)=c}} \beta^{bp},$$

it is sufficient to consider $b$ not divisible by $p$.

If $c$ is in the image of $L(x)$, the assumptions on $L(x)$ guarantee that $L(x) = c$ has $q^d$ distinct roots in $\mathbb{F}_{q^r}$. Let $\zeta$ be a primitive $b$th root of unity in some extension of $\mathbb{F}_{q^r}$. Then $\prod_{i=1}^{b}(L(\zeta^i x) - c)$ is a constant multiple of

$$\begin{aligned}\prod_{i=1}^{b} \prod_{\substack{\beta\in\mathbb{F}_{q^r}\\ L(\beta)=c}} (x - \zeta^{-i}\beta) &= \prod_{\substack{\beta\in\mathbb{F}_{q^r}\\ L(\beta)=c}} \prod_{i=1}^{b}(x - \zeta^{-i}\beta)\\ &= \prod_{\substack{\beta\in\mathbb{F}_{q^r}\\ L(\beta)=c}} (x^b - \beta^b).\end{aligned}$$

From the last expression, the sum we seek is the negative of the coefficient of $x^{b(q^d-1)}$. From the original expression, any nonzero term has degree a sum of $b$ integers in $S = \{0\} \cup \{1, q, q^2, \ldots, q^d\}$. In order to have a nonmultiple of $q$—which $b(q^d - 1)$ is—one of the summands must be 1. We claim that for $b$ in the specified range, $(b-2)$ of the other summands are $q^d$, and the final summand is $q^l$ for $l \in \{0, \ldots, d\}$. We first show that the largest possible term using only $(b-3)$ summands of $q^d$ is not large enough by showing $(b-3)q^d+1+2q^{d-1} < b(q^d-1)$.

Observe that $0 < q^d - 2q^{d-1} + 1$ so

$$b \leqslant 2(q^d - 1) < 3q^d - 2q^{d-1} - 1$$

therefore

$$\begin{aligned} b(q^d - 1) &= bq^d - b\\ &> bq^d - 3q^d + 2q^{d-1} + 1\\ &= (b - 3)q^d + 2q^{d-1} + 1\end{aligned}$$

as claimed.

Now note that 0 may not be one of the summands from $S$ since $(b-2)q^d+1+0 = b(q^d-1)$ implies $b = 2q^d-1$ which is out of the range specified. This shows that $\sum_{L(\beta)=c} \beta^b$ is independent of $c$ as claimed. We also conclude that this value can be nonzero only when $(b - 2)q^d + 1 + q^l = b(q^d - 1)$ where $l \in \{0, \ldots, d\}$. Solving for $b$ we get $b = 2q^d - q^l - 1$. □

**Lemma 4.7.** *Let* $D = \{0\} \cup \{\eta^{vm} : m \in \{1, \ldots, \frac{q^r-1}{v}\}\}$

$$\sum_{c\in D} \sum_{\substack{\alpha\in\mathbb{F}_{q^r}\\ \alpha^u=c}} \alpha^i = \begin{cases} -1 & \textit{if } \frac{(q^r-1)u}{v} \mid i \textit{ and } i > 0,\\ 0 & \textit{otherwise.}\end{cases}$$

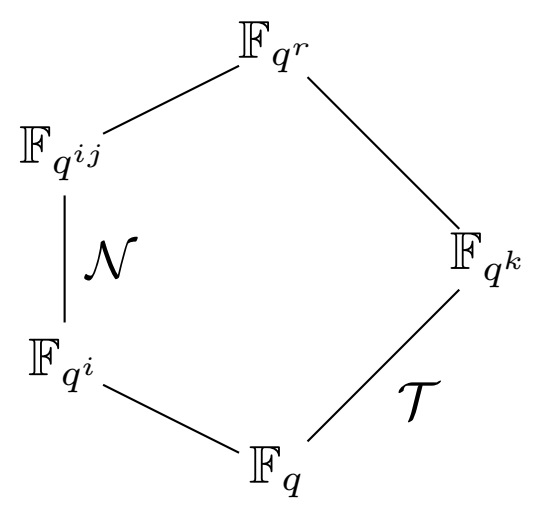

Figure 1: Field extensions defining the curve of Example 4.8

*Proof.* We are summing over all $\alpha$ such that $\alpha = 0$ or $\alpha \in \{\eta^{\frac{v}{u}m} : m \in \{1, \ldots, \frac{(q^r-1)u}{v}\}\}$. Thus for $i = 0$ the sum is $1 + \frac{(q^r-1)u}{v}$. Let $p$ be the prime divisor of $q$ and recall that we assumed that $p \mid (v - u)$. Then $\frac{(q^r-1)u}{v} \equiv -1$ modulo $p$. Therefore $1 + \frac{(q^r-1)u}{v}$ is 0 in $\mathbb{F}_{q^r}$.

For $i > 0$, we may rewrite the sum as

$$\sum_{m=1}^{\frac{(q^r-1)u}{v}} (\eta^{\frac{v}{u}m})^i = \sum_{m=0}^{\frac{(q^r-1)u}{v}-1} (\eta^{\frac{v}{u}i})^m.$$

Now $\eta^{\frac{v}{u}i}$ is a $\frac{(q^r-1)u}{v}$ root of unity. If $i$ is a positive multiple of $\frac{(q^r-1)u}{v}$ then $\eta^{\frac{v}{u}i} = 1$, and the sum is $-1$. Otherwise $\eta^{\frac{v}{u}i}$ is a root of $\sum_{m=0}^{\frac{(q^r-1)u}{v}-1} x^m$ so the sum is zero. □

**Example 4.8.** Let $q$ be a prime power, let $i, j, k$ be positive integers with $j, k > 1$ and let $r$ be a common multiple of $ij$ and $k$. Consider the curve with affine equation

$$x^{\frac{q^{ij}-1}{q^i-1}} = y^{q^{k-1}} + y^{q^{k-2}} + \cdots + y^{q^2} + y^q + y.$$

We may write this as

$$\mathcal{N}_{\mathbb{F}_{q^{ij}}/\mathbb{F}_{q^i}}(x) = \mathcal{T}_{\mathbb{F}_{q^k}/\mathbb{F}_q}(y),$$

which is illustrated in the diagram of fields in Figure 1.

This curve matches the former scheme with $v = \frac{q^r-1}{q-1}$ and $u = \frac{q^{ij}-1}{q^i-1}$. In this case $D = \{0\} \cup \{\eta^{vm} : m \in \{1, \ldots, q-1\}\} = \mathbb{F}_q$, and we can verify that $u$ divides $v$ since $v/u = \frac{q^r-1}{q-1} \cdot \frac{q^i-1}{q^{ij}-1} = \frac{q^r-1}{q^{ij}-1} \cdot \frac{q^i-1}{q-1}$ is an integer. Furthermore, both $u$ and $v$ are 1 modulo $q$ so $q$ divides $v - u$. Thus the assumptions of this section are fulfilled. Because of our choice of $D$, the only points on the curve that we are considering are in $\mathbb{F}_{q^{ij}} \times \mathbb{F}_{q^k}$ and there are $q^{k-1}(1 + (q-1)u)$ of them. There may be other points on the curve, but we have not been able to show that the duality properties of this section can be extended to a larger set of points.

| $q = 2$ | | | |
|---|---|---|---|
| $i, j$ | Norm functions on $x$ | $k$ | Trace functions on $y$ |
| 1,2 | $x^3$ | 2 | $y^2 + y$ |
| 1,4 | $x^{15}$ | 4 | $y^8 + y^4 + y^2 + y$ |
| 1,8 | $x^{255}$ | 8 | $y^{128} + y^{64} + y^{32} + y^{16} + y^8 + y^4 + y^2 + y$ |
| 2,2 | $x^5$ | | |
| 2,4 | $x^{85}$ | | |
| 4,2 | $x^{17}$ | | |

| $q = 4$ | | | |
|---|---|---|---|
| $i, j$ | Norm functions on $x$ | $k$ | Trace functions on $y$ |
| 1,2 | $x^5$ | 2 | $y^4 + y$ |
| 1,4 | $x^{85}$ | 4 | $y^{64} + y^{16} + y^4 + y$ |
| 2,2 | $x^{17}$ | | |

| $q = 16$ | | | |
|---|---|---|---|
| $i, j$ | Norm functions on $x$ | $k$ | Trace functions on $y$ |
| 1,2 | $x^{17}$ | 2 | $y^{16} + y$ |

Table 1: For the field $\mathbb{F}_{256}$ the norm and trace functions appearing in Example 4.8 are listed. The curves considered are obtained for a fixed $q$ by setting any norm function equal to any trace function.

Norm-trace curves as defined in [5] correspond to the curves in this example when $i = 1$ and $j = k = r$, and Hermitian curves correspond to the case when $i = 1$ and $j = k = 2$. Other curves can be obtained which are not Hermitian curves or norm-trace curves as in [5]. In Table 1 we list the possible norm and trace functions for the case $q^r = 2^8$. For instance the curve $x^5 = y^8 + y^4 + y^2 + y$ is obtained taking $q = 2$, $i = 2$, $j = 2$, $k = 4$. In this case $u = 5$ and $v = 15$ and the number of points is 48.

**Remark 4.9.** An alternative approach is to take $D$ to be the powers of $\eta^v$, $D = \{\eta^{vm} : m \in \{1, \dots, \frac{q^r-1}{v}\}\}$, and $\mathcal{M} = \{x^a y^b : a < \frac{(q^r-1)u}{v}, b < q^d\}$. We define for $z = x^a y^b \in \mathcal{M}$,

$$\hat{z} = \begin{cases} y^{q^d-1-b} & \text{when } a = 0, \\ x^{\frac{(q^r-1)u}{v} - a} y^{q^d-1-b} & \text{when } a \neq 0. \end{cases}$$

and for $W \subseteq \mathcal{M}$, $W^\perp = \mathcal{M} \setminus \{\hat{z} : z \in W\}$. Then $C_W = E_{W^\perp}$ provided that for any $x^a y^b \in W$ the monomials $x^a y^{b'}$ are also in $W$ where $b' \in \{b - q^d + q^l : l \in \{0 \dots d\}\} \cap \mathbb{N}_0$.

# 5 Correction-capability-optimized evaluation codes

The rings $A$ in the previous sections are all *order domains* and the codes $C_W$ are what we call order-prescribed (by $W$) codes [2]. In this section, after a brief review of order domains and issues related to decoding performance of the Berlekamp-Massey-Sakata algorithm with majority voting, we will define four types of codes on order domains, differing by the method for choosing $W$. We will see that the $W$ in each of the four code constructions—in the context of the rings $A$ in the previous sections—is divisor closed, so we may easily identify generator matrices for the codes.

## 5.1 Codes from order domains

Given a field $\mathbb{F}$ and an $\mathbb{F}$-algebra $A$, an *order function* on $A$ is a map $\rho : A \longrightarrow \mathbb{N}_{-1}$ which satisfies: i) the set $L_m = \{f \in A : \rho(f) \leqslant m\}$ is an $m+1$ dimensional vector space over $\mathbb{F}$; ii) if $f, g, z \in A$ and $z$ is nonzero then $\rho(f) > \rho(g) \Longrightarrow \rho(zf) > \rho(zg)$ [14, 9, 1]. The pair $A, \rho$ is often called an order domain. It is easy to show that $\rho$ must be surjective.

In each of the previous sections, the ring $A$ admits an order function in which $z_i$ has order $i$. That is, we may define $\rho$ by

**(i)** $\rho(0) = -1$,

**(ii)** $\rho(z_i) = i$,

**(iii)** $\rho(\sum_{i\in I} a_i z_i) = \max I$, where the $a_i's$ are assumed to be nonzero.

We call $\mathcal{B} = \{z_i : i \in \mathbb{N}_0\}$ a *$\rho$-good basis* of $A$. It is a basis for $A$ over $\mathbb{F}$.

An operation $\oplus$ in $\mathbb{N}_0$ can be well defined by $i \oplus j = \rho(fg)$ where $f$ and $g$ are such that $\rho(f) = i$ and $\rho(g) = j$. In fact $\mathbb{N}_0, \oplus$ is a commutative semigroup. We can define a partial ordering $\preccurlyeq$ on $\mathbb{N}_0$ by setting $i \preccurlyeq j$ if and only if there exists $k \in \mathbb{N}_0$ such that $i \oplus k = j$. When $i \preccurlyeq j$ we must also have $i \leqslant j$. An important parameter for decoding is $\nu_i = |\{j \in \mathbb{N}_0 : j \preccurlyeq i\}|$.

## 5.2 Codes designed for prescribed correction capability

As in the previous sections we consider a surjective map $\varphi : A \longrightarrow \mathbb{F}^n$. Given a subset $W$ of $\mathcal{B}$, define the *order-prescribed evaluation code* related to $W$ as the $\mathbb{F}$-subspace $E_W$ generated by $\{\varphi(z_i) : z_i \in W\}$ and define $C_W$ to be its dual code.

The two results on decoding performance that we need are

**Theorem 5.1.** *[4] All error vectors of weight $t$ can be corrected by $C_W$ if $W$ contains all elements $z_i$ with $\nu_i < 2t + 1$.*

**Theorem 5.2.** *[2] All generic error vectors of weight $t$ can be corrected by $C_W$ if $W$ contains all elements $z_i$ with $i \notin \{j \oplus k : j, k \geqslant t\}$.*

The four families of codes we consider are below. The first two correct all errors of a given weight $t$, while the latter two correct all generic errors of weight $t$. The "improved" codes take full advantage of majority voting as discovered by Feng and Rao [4].

**Standard evaluation codes** To design a standard evaluation code which will correct $t$ errors, let $m(t) = \max\{i \in \mathbb{N}_0 : \nu_i < 2t+1\}$. Let $R(t) = \{z_i : i \leqslant m(t)\}$ and $r(t) = |R(t)|$. The code $C_{R(t)}$ has minimum distance at least $2t+1$. Its real redundancy will be given by the number of checks corresponding to $R_\varphi(t) = \{z_i \in R(t) : C_i \neq C_{i-1}\}$.

**Feng-Rao improved codes** To design an order-prescribed evaluation code correcting $t$ errors, we take $\widetilde{R}(t) = \{z_i \in \mathbb{N}_0 : \nu_i < 2t+1\}$ and use the code $C_{\widetilde{R}(t)}$. Let $\widetilde{r}(t) = |\widetilde{R}(t)|$. The Feng-Rao improved code correcting $t$ errors requires $r(t) - \widetilde{r}(t)$ fewer check symbols than the standard code correcting $t$ errors. Again the real redundancy of these codes will be given by the number of checks corresponding to $\widetilde{R}_\varphi(t) = \{z_i \in \widetilde{R}(t) : C_i \neq C_{i-1}\}$.

**Standard generic evaluation codes** To design a standard evaluation code that will correct all generic errors of weight at most $t$, let $m^*(t) = \max(\mathbb{N}_0 \setminus \{i \oplus j : i, j \geqslant t\})$ and let $R^*(t) = \{z_i \in \mathbb{N}_0 : i \leqslant m^*(t)\}$. The number of check symbols for the code $C_{R^*(t)}$ is $r^*(t) = |R^*(t)|$. Its real redundancy will be given by the number of checks corresponding to $R^*_\varphi(t) = \{z_i \in R^*(t) : C_i \neq C_{i-1}\}$.

**Improved generic evaluation codes** To design an order-prescribed evaluation code correcting $t$ generic errors, we use the code $C_{\widetilde{R}^*(t)}$ where $\widetilde{R}^*(t)$ is $\mathbb{N}_0 \setminus \{i \oplus j : i, j \geqslant t\}$. Let $\widetilde{r}^*(t) = |\widetilde{R}^*(t)|$. Clearly $\widetilde{r}^*(t) \leqslant r^*(t)$. The real redundancy of $C_{\widetilde{R}^*(t)}$ will be given by the number of checks corresponding to $\widetilde{R}^*_\varphi(t) = \{z_r \in \widetilde{R}^*(t) : C_r \neq C_{r-1}\}$.

**Example 5.3.** For the Reed-Muller codes over $\mathbb{F}_q[x, y]$, the parity checks $\varphi(x^a y^b)$ can be represented by the corresponding monomials in the $\mathbb{N}_0 \times \mathbb{N}_0$ grid as in Figure 2. In this case, if $z_i = x^a y^b$, then

$$\begin{aligned} \nu_i &= \#\{x^{a'} y^{b'} : x^{a'} y^{b'} \mid x^a y^b\} \\ &= (a+1)(b+1). \end{aligned}$$

In Figure 3 we illustrated the $\nu$-values of the first monomials, according to the representation in Figure 2. Suppose we want to correct 5 errors. Figure 4 represents the parity checks in $R(5)$, $\widetilde{R}(5)$, $R^*(5)$ and $\widetilde{R}^*(5)$.

**Example 5.4.** Consider the Hermitian codes over $\mathbb{F}_{4^2}$. The parity checks $\varphi(x^i y^j)$ can be represented by the corresponding monomials in the $\mathbb{N}_0 \times \mathbb{N}_0$ grid as in Figure 5. Figure 6(a) represents the $(q, q+1)$ graded degree. Figure 6(b) represents the $\nu$-values of the basis monomials. Suppose we want to correct 2 errors. Figure 7 represents the parity checks in $R(2)$, $\widetilde{R}(2)$, $R^*(2)$ and $\widetilde{R}^*(2)$.

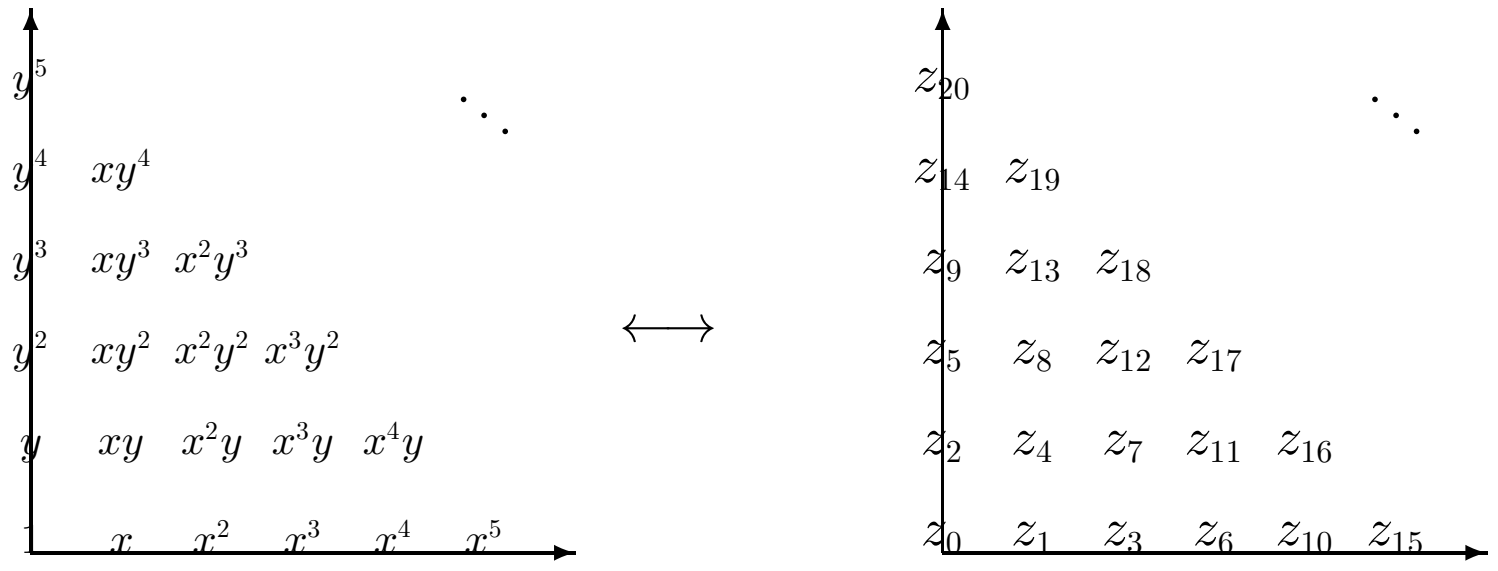

Figure 2: Representation of monomials in the $\mathbb{N}_0 \times \mathbb{N}_0$ grid for Reed-Muller codes.

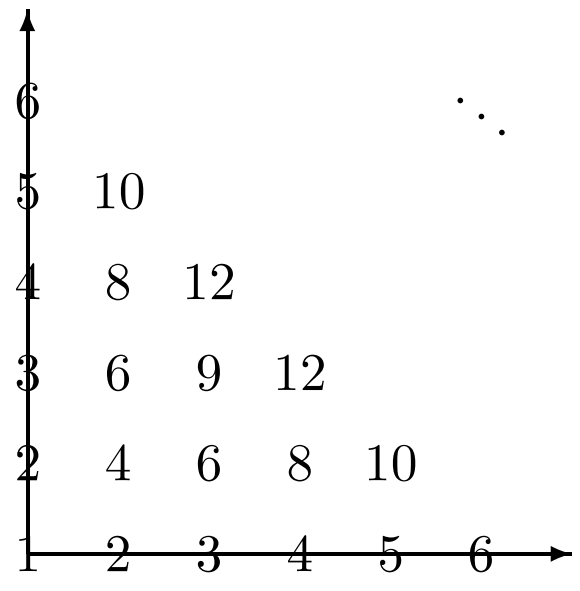

Figure 3: $\nu$-values in the $\mathbb{N}_0 \times \mathbb{N}_0$ grid of the corresponding monomials.

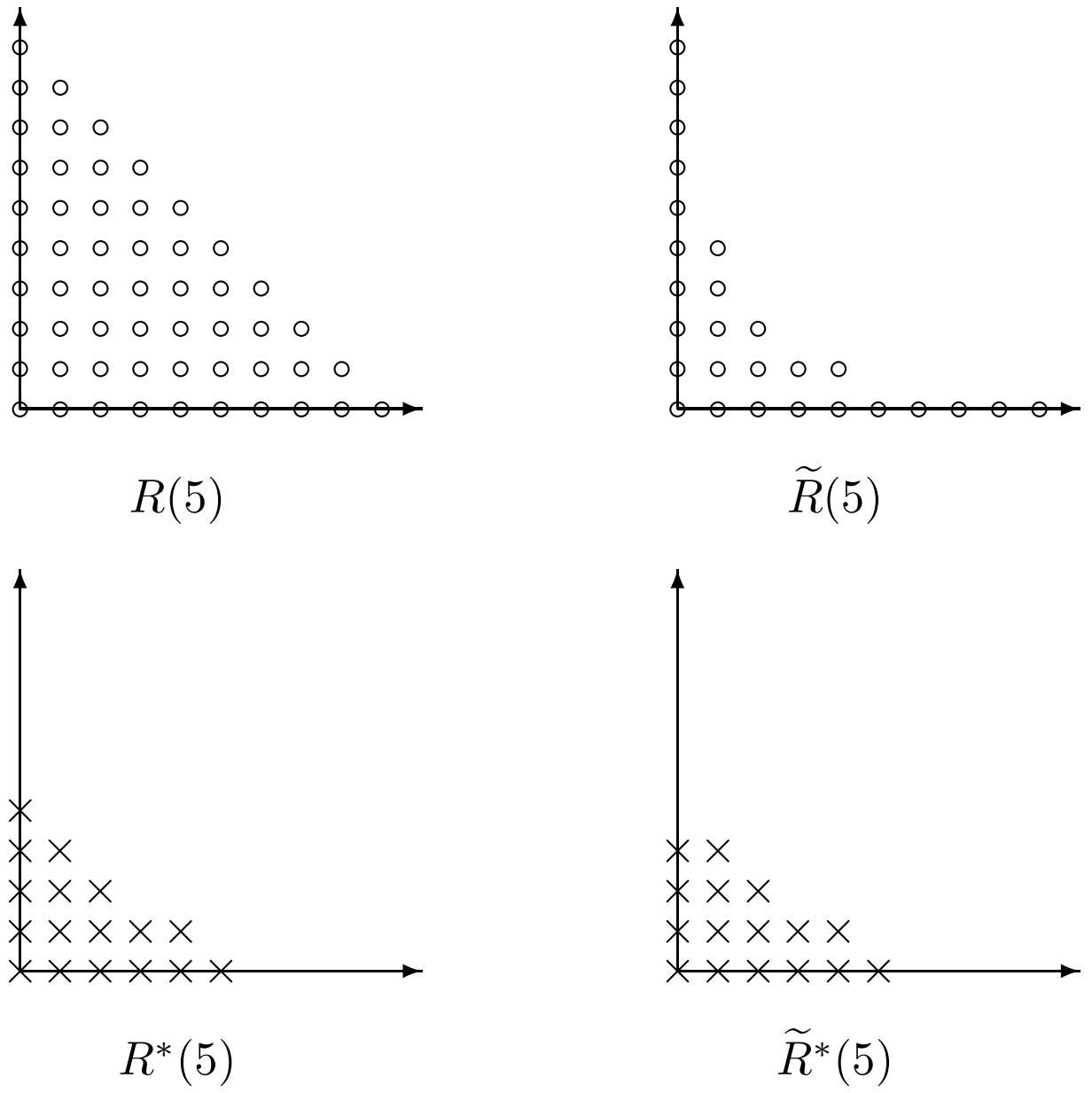

Figure 4: Parity checks required for decoding 5 errors with Reed-Muller codes.

**Example 5.5.** For the codes over the curve with affine equation

$$x^5 = y^8 + y^4 + y^2 + y$$

over $\mathbb{F}_{16}$, the parity checks $\varphi(x^i y^j)$ can be represented by the corresponding monomials in the $\mathbb{N}_0 \times \mathbb{N}_0$ grid as in Figure 8. Figure 9(a) represents the $(q^d, u) = (8, 5)$ graded degree. Figure 9(b) represents the $\nu$-values of the basis monomials. Suppose we want to correct 3 errors. Figure 10 represents the parity checks in $R(3)$, $\widetilde{R}(3)$, $R^*(3)$ and $\widetilde{R}^*(3)$, respectively.

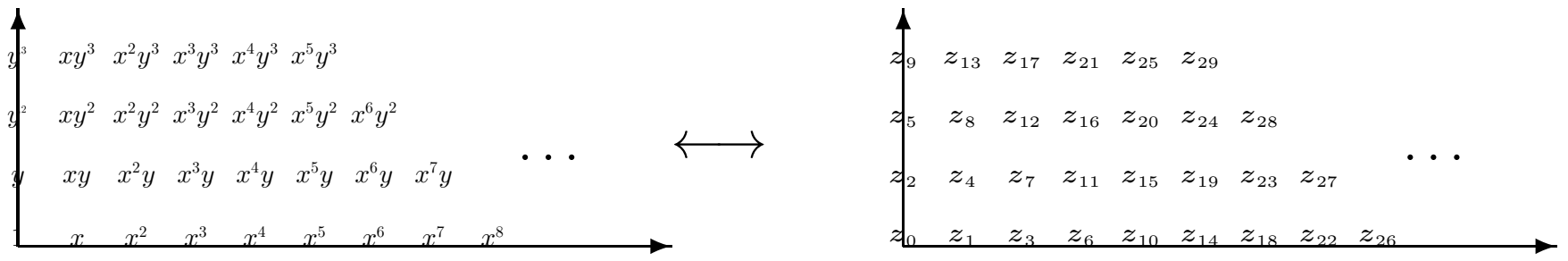

Figure 5: Representation of monomials in the $\mathbb{N}_0 \times \mathbb{N}_0$ grid for the Hermitian codes over $\mathbb{F}_{4^2}$.

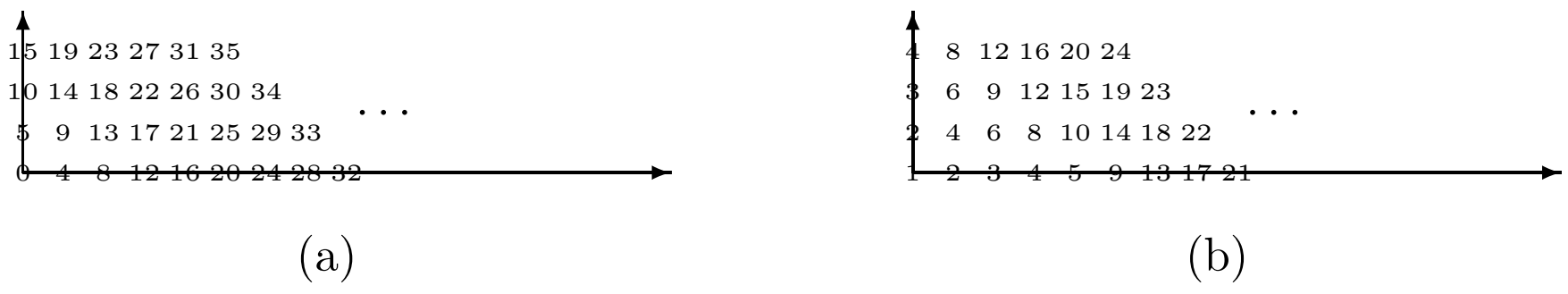

Figure 6: (a) $(q, q+1) = (4,5)$ graded degree, (b) $\nu$-values for the Hermitian codes over $\mathbb{F}_{4^2}$.

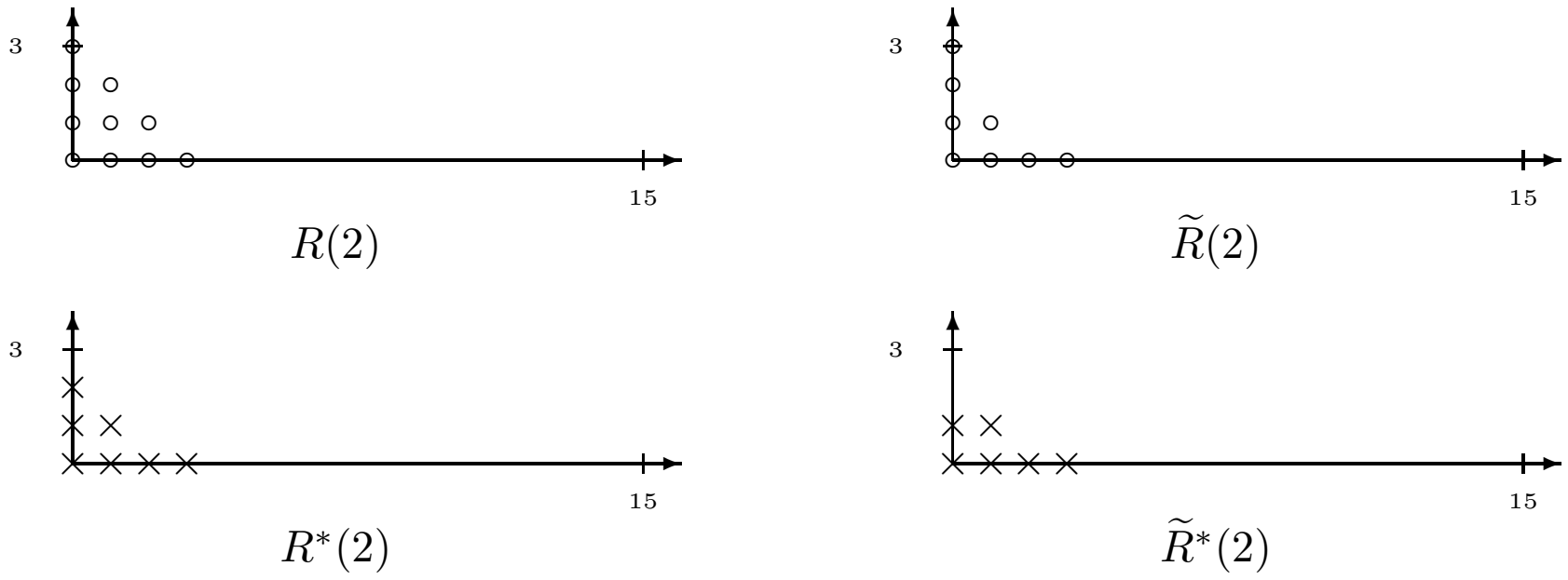

Figure 7: Parity checks required for decoding 2 errors with the Hermitian codes over $\mathbb{F}_{4^2}$.

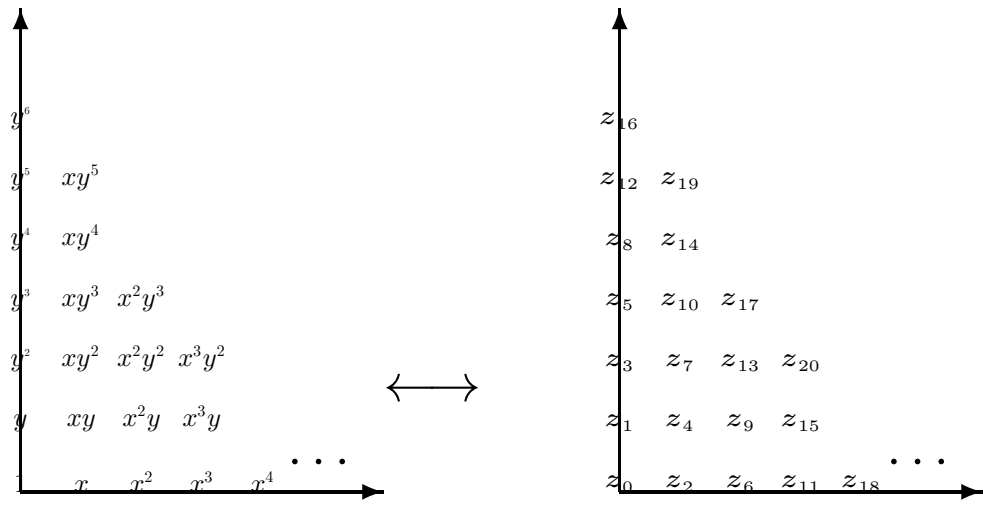

Figure 8: Representation of monomials in the $\mathbb{N}_0 \times \mathbb{N}_0$ grid for the codes over the curve $x^5 = y^8 + y^4 + y^2 + y$.

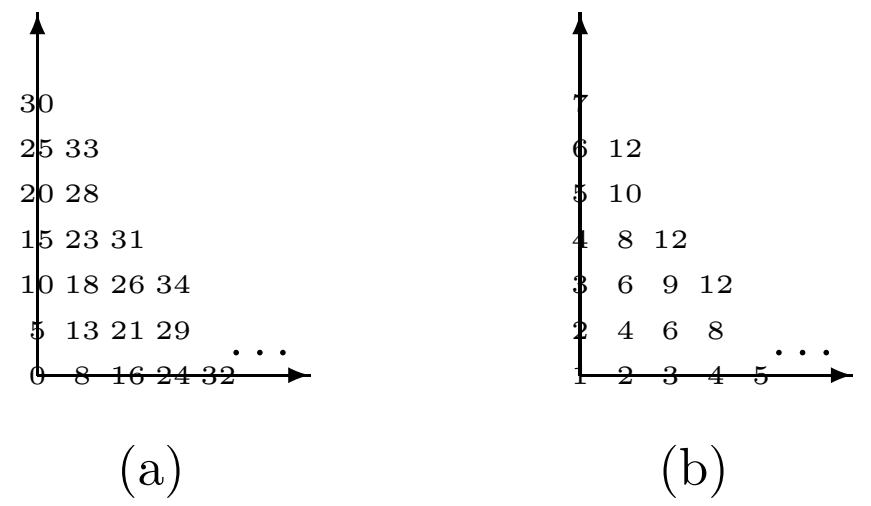

Figure 9: (a) $(q^d, u) = (8, 5)$ graded degree, (b) $\nu$-values for the codes over the curve $x^5 = y^8 + y^4 + y^2 + y$.

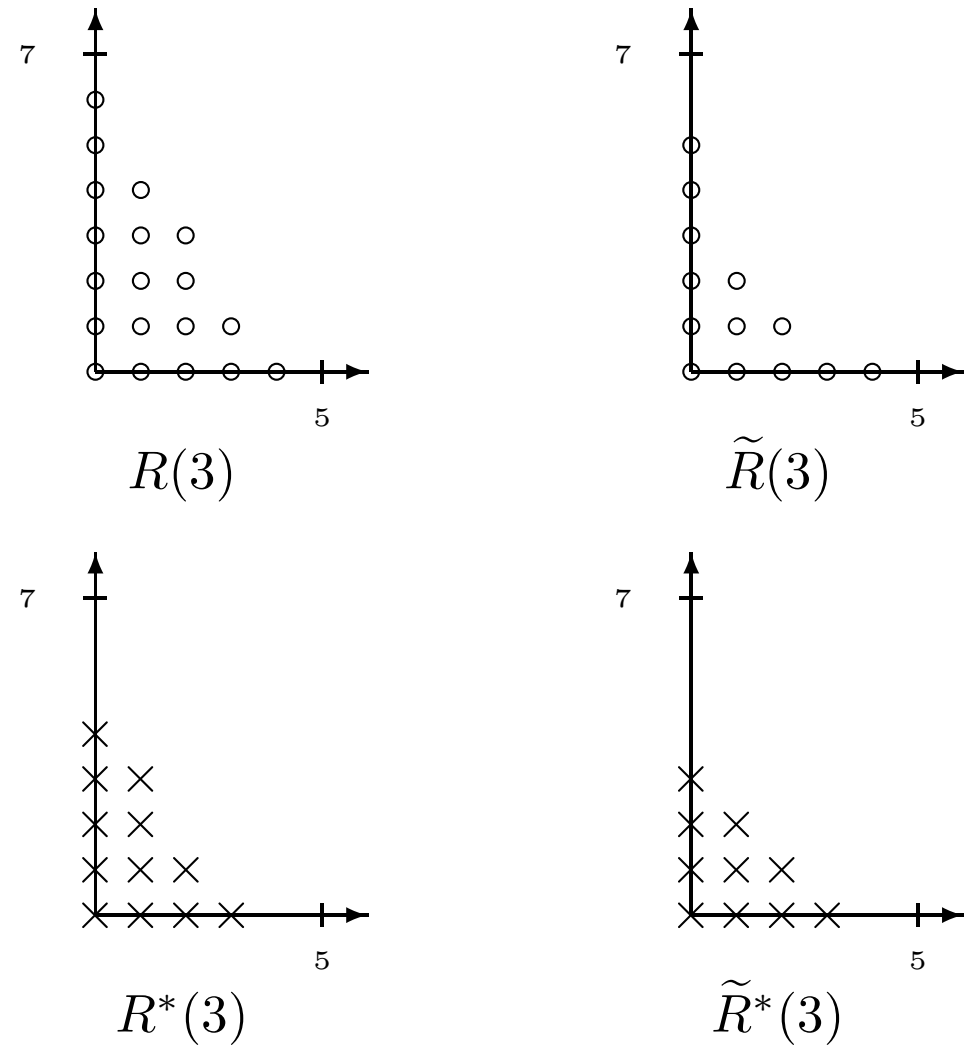

Figure 10: Parity checks required for decoding 3 errors with the codes over the curve $x^5 = y^8 + y^4 + y^2 + y$.

## 5.3 Dimension and generating matrices

We say that a subset $W$ of $\mathcal{B}$ is *closed under* $\preccurlyeq$ when $z_j \in W$ and $i \preccurlyeq j$ implies $z_i \in W$. For the rings $A$ and bases $\mathcal{B}$ of the previous sections, being closed under $\preccurlyeq$ implies being divisor-closed.

**Lemma 5.6.** *The sets $R(t)$, $\widetilde{R}(t)$, $R^*(t)$, $\widetilde{R}^*(t)$ are closed under $\preccurlyeq$.*

*Proof.* The set $R(t)$ is closed under $\preccurlyeq$ because $i \preccurlyeq j$ implies $i \leqslant j$ and if $j \in R(t)$ and $i \leqslant j$ tben $i \in R(t)$. By the same argument, $R^*(t)$ is closed under $\preccurlyeq$.

If $i \preccurlyeq j$ and $j \in \widetilde{R}(t)$ then $\nu_i \leqslant \nu_j < 2t+1$, so $i \in \widetilde{R}(t)$. Thus $\widetilde{R}(t)$ is closed under $\preccurlyeq$.

Finally, to prove that $\widetilde{R}^*(t)$ is closed under $\preccurlyeq$ notice that, if $i \preccurlyeq j$ then $j = i \oplus s$ for some $s \in \mathbb{N}_0$. Suppose $i \notin \tilde{R}^*(t)$. Then $i = k \oplus l$ with $k, l \geqslant t$ and $j = i \oplus s = k \oplus (l \oplus s)$ and so $j \notin \tilde{R}^*(t)$. □

**Corollary 5.7.** *In the context of Reed-Muller codes, toric codes and the codes in Section 4,*

$$\begin{aligned} R_\varphi(t) &= R(t) \cap \mathcal{M}, \\ \tilde{R}_\varphi(t) &= \tilde{R}(t) \cap \mathcal{M}, \\ R^*_\varphi(t) &= R^*(t) \cap \mathcal{M}, \\ \tilde{R}^*_\varphi(t) &= \tilde{R}^*(t) \cap \mathcal{M}. \end{aligned}$$

**Corollary 5.8.** *In the context of Reed-Muller codes, toric codes and the codes in Section 4,*

$$\begin{aligned} C_{R(t)} &= C_{R_\varphi(t)} = & E_{R_\varphi(t)^\perp}, \\ C_{\widetilde{R}(t)} &= C_{\widetilde{R}_\varphi(t)} = & E_{\tilde{R}_\varphi(t)^\perp}, \\ C_{R^*(t)} &= C_{R^*_\varphi(t)} = & E_{R^*_\varphi(t)^\perp}, \\ C_{\widetilde{R^*}(t)} &= C_{\widetilde{R^*}_\varphi(t)} = & E_{\tilde{R}^*_\varphi(t)^\perp}. \end{aligned}$$

**Example 5.9.** For Reed-Muller codes, the minimal set of parity checks depends on the field over which the codes are defined. For Reed-Muller codes over $\mathbb{F}_8[x,y]$ correcting 5 errors, $R^*(5)$ and $\widetilde{R}^*(5)$ are included in $\mathcal{M}$ and thus $R^*_\varphi(5) = R^*(5)$ and $\widetilde{R}^*_\varphi(5) = \widetilde{R}^*(5)$, but $R(5)$ and $\widetilde{R}(5)$ are not included in $\mathcal{M}$ and so $R_\varphi(5) \neq R(5)$ and $\widetilde{R}_\varphi(5) \neq \widetilde{R}(5)$. However if we consider Reed-Muller codes over $\mathbb{F}_4[x,y]$ correcting 5 errors, then none of the sets $R(5), \widetilde{R}(5), R^*(5), \widetilde{R}^*(5)$ is included in $\mathcal{M}$ and we can drop redundant parity checks in all cases. In Figure 11 and Figure 12 we illustrated the sets $\widetilde{R}(5)$, $\widetilde{R}^*(5)$, $\widetilde{R}_\varphi(5)$, $\widetilde{R}^*_\varphi(5)$, $\widetilde{R}_\varphi(5)^\perp$, $\widetilde{R}^*_\varphi(5)^\perp$, for the codes over $\mathbb{F}_8$ and $\mathbb{F}_4$, respectively. Notice that for the codes over $\mathbb{F}_4$, $\widetilde{R}_\varphi(5) = \widetilde{R}^*_\varphi(5)$. For the codes $C_{R(5)}$ and $C_{R^*(5)}$ the construction is analogous.

**Example 5.10.** Consider the codes over the Hermitian curve over $\mathbb{F}_{4^2}$ correcting 2 errors, as in Figure 7. In this case, $\frac{(q^r-1)u}{v} = 15$ and $q^d - 1 = 3$. Since all checks are inside $\mathcal{M}$ we can not supress any. In Figure 13 we represented how the set of generating vectors is obtained from the set of parity checks. We just illustrated the codes $C_{\widetilde{R}(2)}$ and $C_{\widetilde{R}^*(2)}$. For the codes $C_{R(2)}$ and $C_{R^*(2)}$ the construction is analogous.

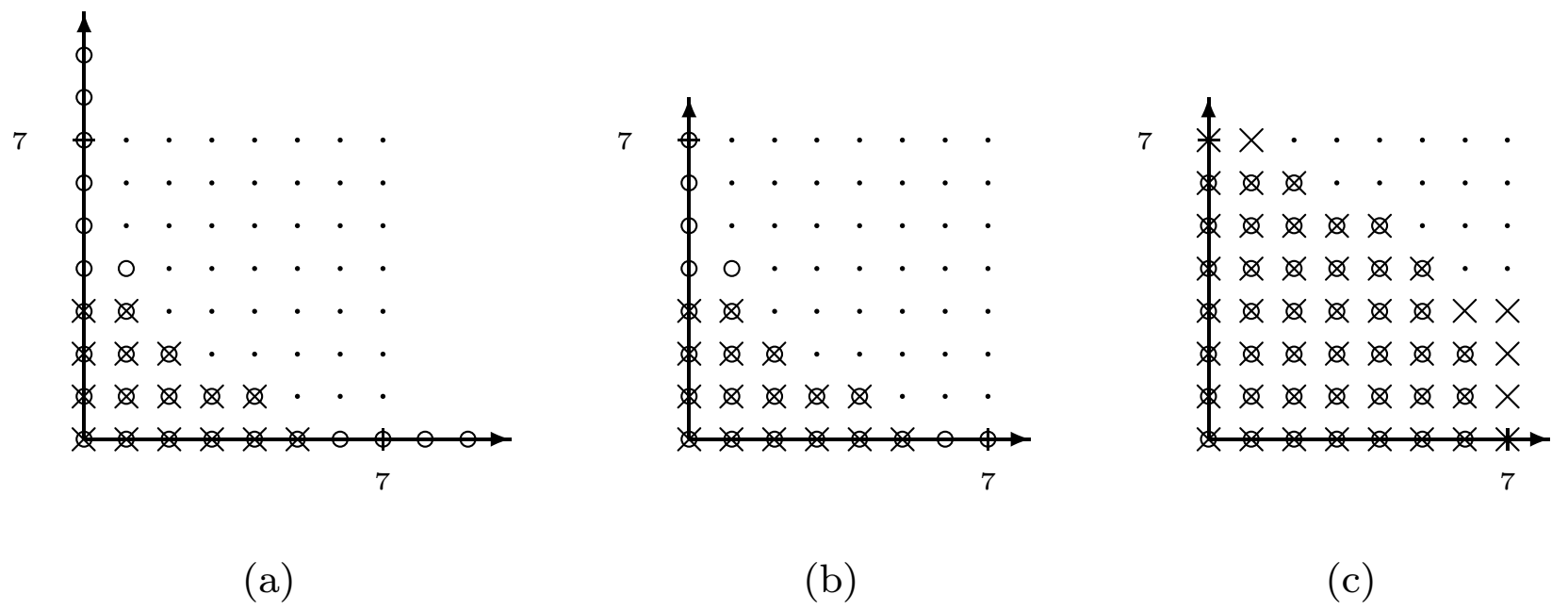

Figure 11: Reed-Muller codes over $\mathbb{F}_8[x,y]$: (a) shows $\widetilde{R}(5)$ (with $\circ$) and $\widetilde{R}^*(5)$ (with $\times$); (b) shows $\widetilde{R}_\varphi(5)$ and $\widetilde{R}^*_\varphi(5)$; (c) shows $\widetilde{R}_\varphi(5)^\perp$ and $\widetilde{R}^*_\varphi(5)^\perp$.

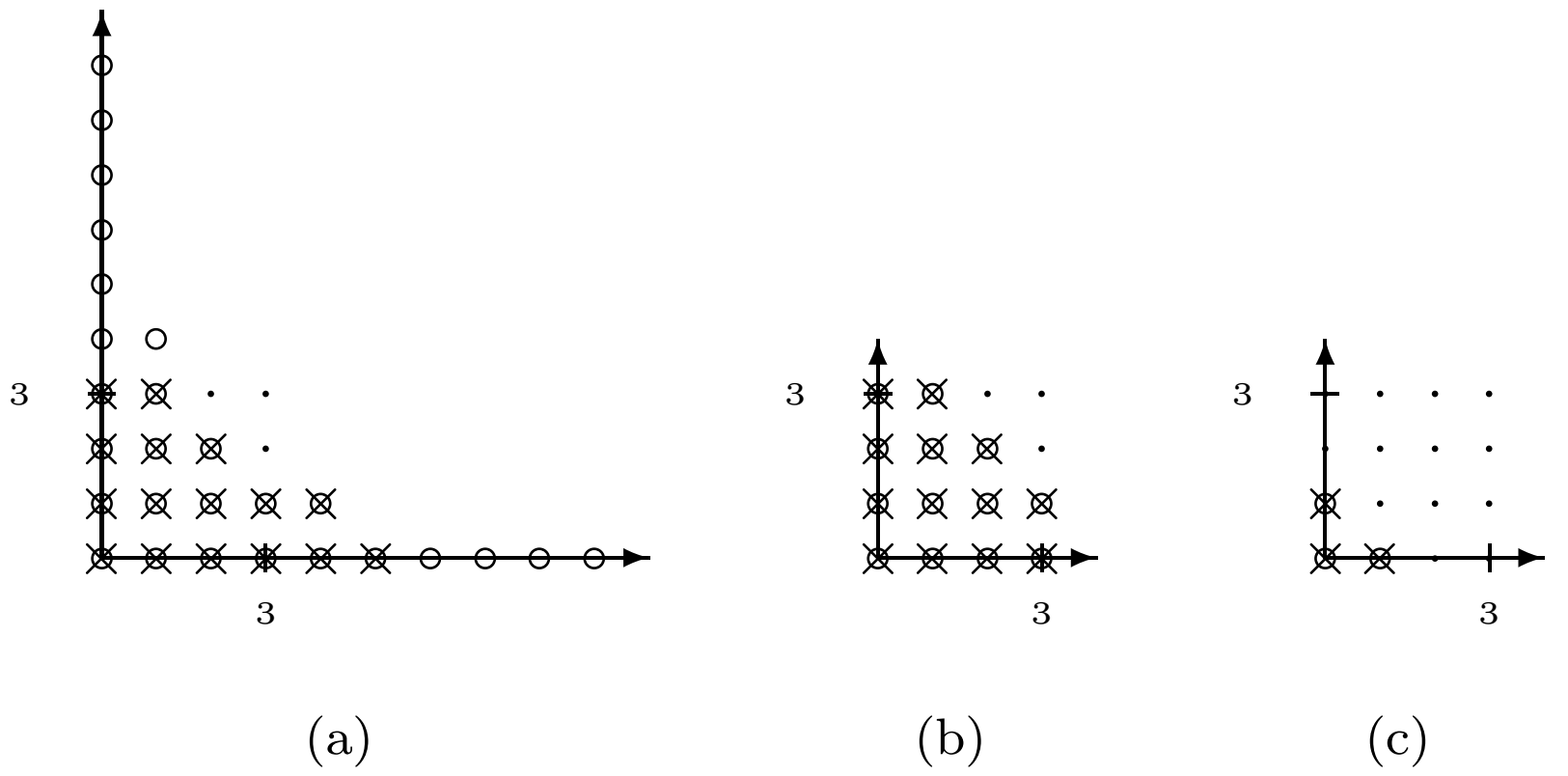

Figure 12: Reed-Muller codes over $\mathbb{F}_4[x,y]$: (a) shows $\widetilde{R}(5)$ (with $\circ$) and $\widetilde{R}^*(5)$ (with $\times$); (b) shows $\widetilde{R}_\varphi(5)$ and $\widetilde{R}^*_\varphi(5)$; (c) shows $\widetilde{R}_\varphi(5)^\perp$ and $\widetilde{R}^*_\varphi(5)^\perp$.

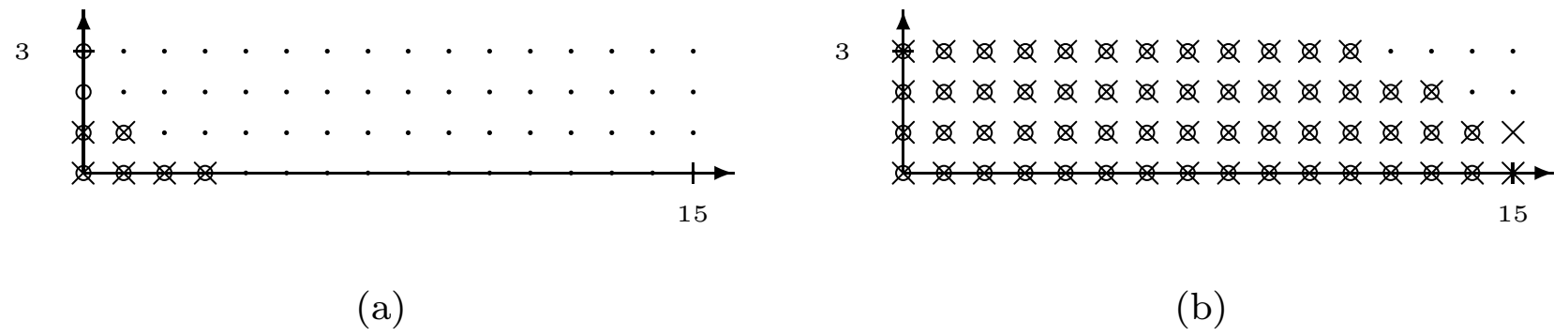

Figure 13: Hermitian codes over $\mathbb{F}_{4^2}$: (a) shows $\widetilde{R}(2) = \widetilde{R}_{\varphi}(2)$ (with $\circ$) and $\widetilde{R}^*(2) = \widetilde{R}^*_{\varphi}(2)$ (with $\times$); (b) shows $\widetilde{R}_{\varphi}(2)^{\perp}$ and $\widetilde{R}^*_{\varphi}(2)^{\perp}$.

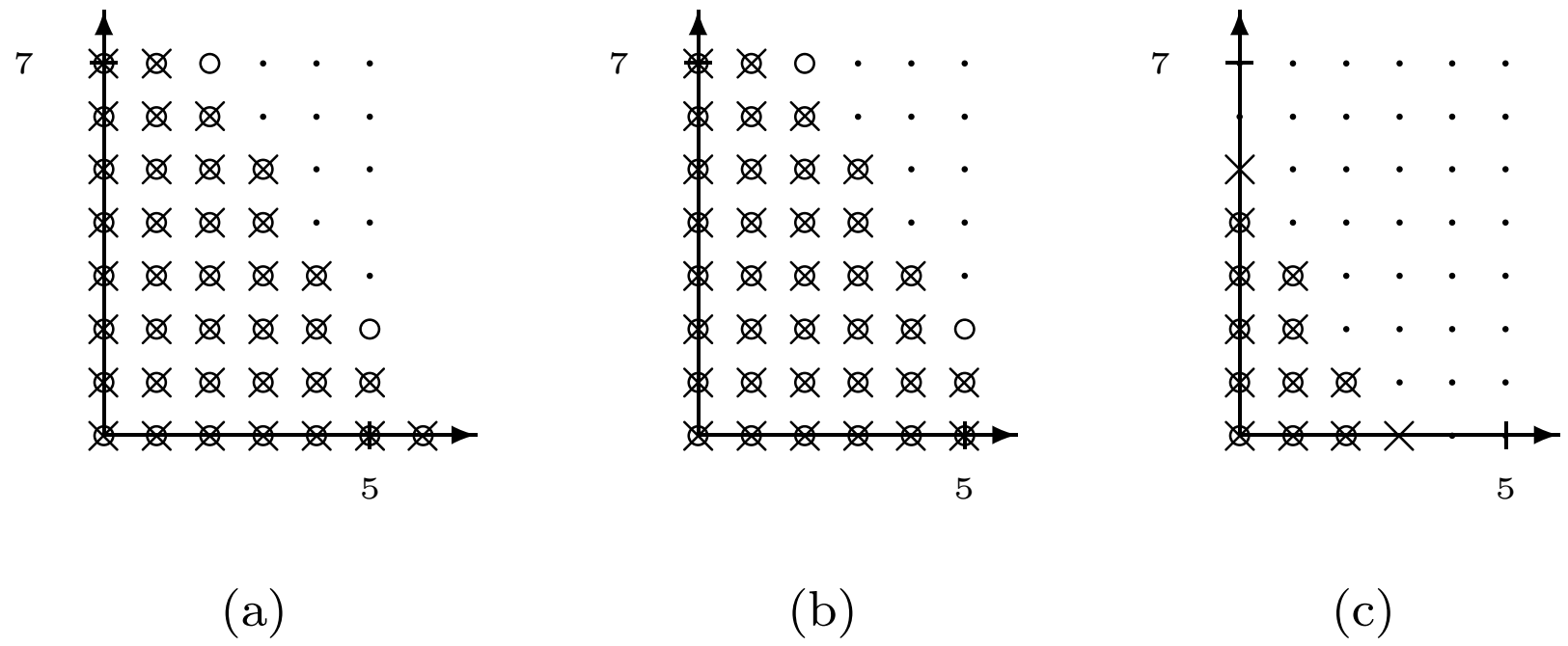

Figure 14: Codes from the affine curve $x^5 = y^8 + y^4 + y^2 + y$: (a) shows $\widetilde{R}(12)$ (with $\circ$) and $\widetilde{R}^*(12)$ (with $\times$); (b) shows $\widetilde{R}_{\varphi}(12)$ and $\widetilde{R}^*_{\varphi}(12)$; (c) shows $\widetilde{R}_{\varphi}(12)^{\perp}$ and $\widetilde{R}^*_{\varphi}(12)^{\perp}$.

**Example 5.11.** Consider the codes with affine curve $x^5 = y^8 + y^4 + y^2 + y$ correcting 12 errors. We have $\frac{(q^r-1)u}{v} = 5$ and $q^d - 1 = 7$. In this case $R(12) = \widetilde{R}(12)$ and $R^*(12) = \widetilde{R}^*(12)$. In Figure 14 we represented how redundant parity checks are dropped for both cases and how the set of generating vectors is obtained from the set of parity checks.